\newcommand\g{\left(1 - {2 G M\over r}\right)} 
\newcommand\eV{{\mathord{\hbox{eV}}}}
\newcommand\MeV{{\mathord{\hbox{MeV}}}}
\newcommand\local{{\mathord{\hbox{\scriptsize{local}}}}}
\documentstyle[12pt]{article} 
\begin{document} 
\pagestyle{empty}

\centerline{\large{\bf Comment on}} 

\vspace{.5cm} 

\centerline{\large{\bf ``Gravitationally Induced Neutrino-Oscillation
Phases''}} 

\vspace{2cm}

\centerline{\bf Tanmoy Bhattacharya$^1$, Salman Habib$^2$, and Emil
Mottola$^1$} 

\vspace{1cm}

\centerline{\em $^1$Theoretical Division}
\centerline{\em T-8, MS B285}
\centerline{\em Los Alamos National Laboratory}
\centerline{\em Los Alamos, NM 87545}

\vspace{.5cm}

\centerline{\em $^2$Theoretical Division}
\centerline{\em T-DOT, MS B288}
\centerline{\em Los Alamos National Laboratory}
\centerline{\em Los Alamos, NM 87545}

\vspace{2cm}

\centerline{\bf Abstract}

\vspace{.2cm}

We critically examine the recent claim of a ``new effect'' of
gravitationally induced quantum mechanical phases in neutrino
oscillations. A straightforward exercise in the Schwarzschild
coordinates appropriate to a spherically symmetric non-rotating star
shows that, although there is a general relativistic effect of the
star's gravity on neutrino oscillations, it is not of the form
claimed, and is too small to be measured.

\newpage
\pagestyle{plain}
\pagenumbering{arabic}

In a recent communication \cite{Ahluwalia}, Ahluwalia and Burgard
claim to have discovered a ``new effect from an hitherto unexplored
interplay of gravitation and the principle of the linear superposition
of quantum mechanics.'' In fact, the calculation of the quantum
mechanical phase of a particle propagating in the geometry of a
collapsed star appears in several textbooks on general relativity
\cite{MTW}. More importantly, the claimed results in equations (6) to
(8) of Ref. \cite{Ahluwalia} appear to be at variance with the
standard treatment, which we review in this Comment. The gravitational
correction to the phase turns out to be of order $10^{-9}$ for
neutrinos in the eV mass range and is therefore completely negligible
for situations of astrophysical interest.

In the geometrical optics limit the quantum mechanical phase
accumulated by a particle propagating from point A to point B in the
gravitational field described by the metric $g_{\mu\nu}$ is given by
the classical action of the particle, divided by $\hbar$, namely,
\begin{equation}
\Phi = {1\over \hbar} \int_A^B \, m\,ds = {1\over \hbar}\int_A^B p_\mu
dx^\mu = {1\over \hbar}\int_A^B(-E\,dt + p_i\,dx^i)
\label{eq:phi}  
\end{equation}
where $p_\mu$ is the four momentum conjugate to $x^\mu$:
\begin{equation}
p_\mu = m \, g_{\mu\nu}\, {dx^\nu\over ds}~,         \label{eq:pmu}
\end{equation}
and $ds$ is an element of proper length of the particle's wordline.
The integrand of Eqn. (\ref{eq:phi}) is obviously an invariant under
coordinate transformations. However, the form of the integral will
depend on the labeling of the end-points A and B. Equation
(\ref{eq:phi}) is the same as the Eqn. (4) of Ref. \cite{Ahluwalia}
with which those authors begin.

The authors of Ref. \cite{Ahluwalia} address the radial propagation of
relativistic neutrinos in the potential of a spherically symmetric
non-rotating star which is described by the Schwarzschild line element
\begin{equation}
ds^2 = -\g dt^2 + \g^{-1} dr^2 + r^2 (d \theta^2 + \sin^2 \theta\,
d\phi^2)~.  
\end{equation}
We note that the semiclassical phase for radial motion in a
spherically symmetric background does not depend on the spin of the
particle, as can be verified by explicit calculation using the spin
connection in the Dirac equation in this background \cite{Brill}.
Hence Eqn. (\ref{eq:phi}) applies equally well to neutrinos as to
scalar particles in the case of radial motion.  Because the spacetime
has a time-like Killing vector, $\partial/\partial t$, the momentum
conjugate to it is time independent:
\begin{equation}
E \equiv -p_t = m \,\g\, {dt \over ds} = \hbox{constant}~.
\label{eq:E}
\end{equation}
The value of this constant $E$ is the asymptotic energy of the
neutrino at $r = \infty$. For radial motion, the mass shell constraint
reads
\begin{equation}
g^{\mu\nu}p_\mu p_\nu + m^2 = 0 =-\g^{-1} E^2 + \g p_r^2 + m^2~,
\end{equation}
from which we obtain
\begin{equation}
p_r \g  =  \sqrt{ E^2 - m^2 + { 2 G M m^2\over r} }~.
\label{pr} 
\end{equation}
Making use of the definitions (\ref{eq:pmu}) 
\begin{equation}
p_r =  m\, \g^{-1}\, {dr \over ds}~,
\end{equation}
and (\ref{eq:E}), we can write
\begin{equation}
{dt \over dr} =  \g^{-2}\, {E \over p_r}~.
\label{dtdr}
\end{equation}
We regard the Schwarzschild radial coordinates $r_A$ and $r_B$ as
fixed and express the phase $\Phi$ in those coordinates as follows:
\begin{eqnarray}
\Phi &=& {1\over\hbar}\int_{r_A}^{r_B} \left( -E {dt \over dr} + p_r
\right) dr 
          \nonumber\\
     &=& {1\over\hbar} \int_{r_A}^{r_B} \left[ - {E^2\over\g^2} +
     p_r^2 
           \right] {dr\over p_r} \nonumber\\
     &=& -{1\over\hbar}\int_{r_A}^{r_B} { m^2 dr \over \g p_r }
     \nonumber\\ 
     &=& - {m^2 \over \hbar}
           \int_{r_A}^{r_B} {dr \over \sqrt{E^2 - m^2 + { 2 G M
     m^2\over r}}}~. 
\label{phians}
\end{eqnarray}
This is just the standard result \cite{MTW}.  In the weak field
approximation, we expand this result to first order in $G$, to obtain
\begin{equation}
\Phi \simeq -{m^2 \over \hbar\sqrt{E^2 - m^2}} (r_B - r_A) +
            {G M m^4 \over \hbar\left(E^2 - m^2\right)^{3/2}}
            \ln \left({r_B \over r_A}\right) + \dots~.
\end{equation}
The precise application of the above formula depends on the physical
situation at hand, specifically, on what variables are to be held fixed
in a particular interference experiment.

Let us consider the case of neutrinos produced at fixed asymptotic
energy $E$ in a weak flavor eigenstate that is a linear superposition
of mass eigenstates, $m_1$ and $m_2$. Since the energy is fixed but
the masses are different, if interference is to be observed at the
same final spacetime point $(r_B,t_B)$, the relevant components of the
wave function could not both have started from the same initial
spacetime point $(r_A,t_A)$ in the semiclassical
approximation. Instead the lighter mass (hence faster moving)
component must either have started at the same time from a spatial
location $r<r_A$, or (what is equivalent) started from the same
location $r_A$ at a later time $t_A + \Delta t$. Hence, there is
already an initial phase difference between the two mass components
due to this time lag, even before the transport from $r_A$ to $r_B$
which leads to the phase $\Phi$ in (\ref{phians}). The additional
initial phase difference may be taken into account most conveniently
from the second point of view, {\it i.e.} by treating the spatial
coordinates $r_A$ and $r_B$ as fixed and the time of transit,
\begin{equation}
\int_A^B dt = \int_{r_A}^{r_B} {dt\over dr} dr
\end{equation}
as the dependent variable through Eqns. (\ref{pr}) and
(\ref{dtdr}). The difference of this time of transit between the two
mass eigenstates, multiplied by $E$ is precisely the additional phase,
$E\Delta t$ which we must add to $\Delta\Phi$ to obtain the correct
relative phase between the two mass components of the wave function
which interfere at $(r_B, t_B)$ with fixed energy $E$. We note that
the COW experiment \cite{COW} may be treated by similar reasoning and
that many other scenarios for neutrino oscillations which may be
envisaged lead to the same result.

Hence we are led to compute instead of $\Phi$, the quantity,
\begin{equation}
\Phi_r=\Phi + E\int_A^B dt~ = \int_A^B p_r\,dr~  \label{phirad}
\end{equation}
This $\Phi_r$ may be calculated just as easily as the full $\Phi$:
\begin{equation}
\Phi_r={1\over\hbar}\int_A^B{\sqrt{E^2-m^2+{2GMm^2\over r}}\over
\left(1-{2GM\over r}\right)}dr~. 
\label{phirad2}
\end{equation}
In the weak field expansion this becomes,
\begin{equation}
\Phi_r\simeq{1\over\hbar}\sqrt{E^2-m^2}\left[(r_B-r_A)
+ 2GM\ln\left({r_B\over r_A}\right)\right]+{GMm^2\over\hbar
\sqrt{E^2-m^2}}\ln\left({r_B\over r_A}\right) + \cdots       
\label{phiradwk}
\end{equation}
The relative phase difference $\Delta \Phi_r$ between two mass
eigenstates of relativistic neutrinos ($E^2 \gg m^2$) with mass
squared difference of $\Delta m^2$ created at point $r_A$ and
interfering at point $r_B$ is then
\begin{equation}
\Delta\Phi_r \simeq {(\Delta m^2)c^3\over 2\hbar E}(r_B-r_A)
+ {(\Delta m^4)c^7\over 4\hbar E^3}(r_B-r_A)-{(\Delta m^4)
c^5\over 2\hbar E^3}GM\ln\left({r_B\over r_A}\right)+\dots~,  
\label{dphi} 
\end{equation}
where $c$ has been restored to facilitate numerical calculations.  We
note that in the relativistic limit this result is precisely minus
half of the equivalent quantity computed from the full phase
$\Phi$. The first term ($\Delta \Phi_r^0$) in Eqn. (\ref{dphi}) is the
standard flat space result, well known in both neutrino and
strangeness oscillations. The leading order correction to this
familiar result has cancelled in the relativistic limit and we are
left only with the latter higher order terms in Eqn. (\ref{dphi}). The
second term is the special relativistic correction to the phase which
is usually neglected for light neutrinos, and the last term is the
effect of the gravitational field of the star in static Schwarzschild
coordinates which enters only at the same higher order in
$1/E^3$. Numerically its magnitude is equal to
\begin{equation}
  3.74 \times 10^{-9} \left(M\over M_\odot\right) \left( \Delta
  m^4\over \eV^4 \right) \left( \MeV \over E \right)^3
  \ln\left({r_B\over 
r_A}\right) 
\label{gravcorr}
\end{equation}
which is completely negligible in typical astrophysical applications. 

We note more generally that for the case of radial motion in
coordinates such that $g_{rt} = 0$, the expression for the phase,
Eqn. (\ref{eq:phi}), can be written as
\begin{eqnarray}
\Phi &=& -{m^2\over \hbar} \int_A^B {dr_{\local} \over p_{\local}}
\nonumber\\
     &=& -{m^2 c^4 \over \hbar} \int_A^B {dt_{\local} \over
     E_{\local}}\nonumber\\ 
     &=& -{m^2 c^4 \over \hbar} T_{AB} \left( 1 \over E_{\local}
     \right)_{av} 
\end{eqnarray}
where
\begin{eqnarray}
dr_{\local} &=& \sqrt{g_{rr}} dr \nonumber\\
dt_{\local} &=& \sqrt{-g_{tt}} dt  \nonumber\\
p_{\local}  &=& p_r / \sqrt{g_{rr}}  \nonumber\\
E_{\local}  &=& E / \sqrt{-g_{tt}} \qquad {\rm and}\nonumber\\
T_{AB} &=& \int_A^B dt_{\local}
\end{eqnarray}
which is a statement of the equivalence principle. Locally there are
no observable effects of the gravitational potential. The redshifted
energy $E_{\local}$ is not a constant of motion and cannot be pulled
outside of integrals, which accounts for the appearance of
$\left(1/E_{\local}\right)_{av} = (1/T_{AB})\int_A^B
dt_{\local}/E_{\local}$ in the above expression.

The authors of Ref. \cite{Ahluwalia} claim to find an effect on the
phase, first order in $G$\footnote{Actually the phase quoted in
Ref. \cite{Ahluwalia} Eqn. (6) differs from the one obtained by us by
a factor of two, even in flat space, because $\Delta\Phi_r$ enters
interference probabilities through cos $(\Delta\Phi_r)$ whereas their
Eqn. (6) contains sin$^2$ of a phase angle which should be
$\Delta\Phi_r/2$ by the half-angle formula. This discrepancy may have
arisen from a confusion between $\Phi$ and $\Phi_r$ which also differ
by a factor of $-{1\over 2}$, as noted previously.}, equal to
(Eqn. (8) of Ref. \cite{Ahluwalia}) 
\begin{equation}
{G M c \over \hbar} \left[\int_A^B {dr \over r}\right] {\Delta m^2
\over E}~. 
\label{claim}
\end{equation}
However no derivation is given to support this claim and their
quantity $E$ is never defined. If $E$ is the constant of the motion
defined in Eqn. (\ref{eq:E}), then this claim disagrees with the
standard result (\ref{dphi}) rederived here, and is therefore
incorrect. On the other hand, if $E$ is to be identified with our
$E_\local$, then it cannot be pulled out of the integral, and
(\ref{claim}) is incorrect for that reason.

There is a sense in which the {\it first} ($\Delta\Phi_r^0$) term of
Eqn. (\ref{dphi}) has a contribution similar in form to
(\ref{claim}). If we define   
\begin{equation}
\bar{E}_{\local}={1\over r_B-r_A}\int_B^A{E\, dr\over\sqrt{1-{2GM\over
r}}}~, 
\end{equation}
then in the weak field limit,
\begin{equation}
\bar{E}_{\local}\simeq E+{EGM\over r_B-r_A}\ln\left({r_B\over
r_A}\right)  
\label{eloc}
\end{equation}
and we can write
\begin{equation}
\Delta\Phi_r^0 \simeq{\Delta m^2c^3\over 2\hbar
\bar{E}_{\local}}(r_B-r_A)+{\Delta m^2 GMc \over
2\bar{E}_{\local}}\ln\left({r_B\over r_A}\right)
\label{phisplit}
\end{equation}
which is of the form reported in Ref. \cite{Ahluwalia}. However, to
interpret this rewriting of the standard result in different variables
as a ``new'' gravitational effect would be very misleading. The
quantity $E_{\local}$ refers to the energy measured by local observers
at fixed $r$ and differs from the asymptotic $E$ precisely because of
the well known gravitational redshift effect of general
relativity. Since all measuring rods and clocks are subject to the
{\it same} redshift, there are no physical consequences of this local
redshift effect on the observable physics of neutrino oscillations
(for example, if along with the local energy one uses the proper,
rather than the coordinate, length, then the ``effect'' in
(\ref{phisplit}) disappears). In particular, any effect(s) of neutrino
oscillations on energy transport and heating in supernova explosions
are quite indifferent to such local redefinitions of length, time and
energy scales. Provided all calculations are done in a
relativistically covariant framework, local redshift effects are
accounted for automatically and there are no observable consequences
for supernova evolution to be deduced from the decomposition in
(\ref{phisplit}). Of course, if one does {\it not} use a
relativistically covariant framework in the calculations, the error
made will be precisely of the order of the second term in
(\ref{phisplit}).

The only indication of the basis for the claim in
Ref. \cite{Ahluwalia} is a reference to a paper by Stodolsky
\cite{Stodolsky}. However, as Stodolsky himself notes, the split
between ``flat'' and ``curved'' space effects in equation (2.3) of his
paper is coordinate dependent. Hence there is no invariant meaning to
the splitting of the phase into these two pieces, and such a splitting
is completely misleading for the present application, just as is the
splitting of $\Delta \Phi_r^0$ into the two pieces in (\ref{phisplit})
above. In addition, the time component of the quantity Stodolsky calls
the ``usual four-momentum of special relativity'' is not a constant of
motion in the present application and cannot be removed from integrals
over $r$. Since Stodolsky starts with precisely the same phase $\Phi$
of Eqn. (\ref{eq:phi}) the {\it sum} of his two pieces is precisely
equal to the same result (\ref{phians}) rederived here, as may be
checked directly from the definitions in Ref. \cite{Stodolsky}.

Finally we note that the gravitational effect which we have computed
here in (\ref{gravcorr}) has a different dependence on the neutrino
masses and energy from the flat space result, $\Delta\Phi_r^0$, and
hence it cannot be absorbed into $\Delta\Phi_r^0$ by a coordinate
transformation.

\end{document}